\documentstyle[12pt,equation]{article}
\advance\voffset by 28 pt
\begin{document}
\newcommand{\ccbar}{\mbox{$c \overline{c} $}}
\newcommand{\Y}{\mbox{$\Upsilon$}}
\newcommand{\aem}{\mbox{$\alpha_{{\rm em}}$}}
\newcommand{\rmcosh}{\mbox{\rm cosh}}
\newcommand{\rmln}{\mbox{\rm ln}}
\newcommand{\gev}{\mbox{\rm GeV}}
\newcommand{\coss}{\mbox{\rm cos}}
\newcommand{\sine}{\mbox{\rm sin}}
\newcommand{\shat}{\mbox{$\hat{s}$}}
\newcommand{\that}{\mbox{$\hat{t}$}}
\newcommand{\uhat}{\mbox{$\hat{u}$}}
\newcommand{\sighat}{\mbox{$\hat{\sigma}$}}
\newcommand{\rs}{\mbox{$\sqrt{s}$}}
\newcommand{\pT}{\mbox{$p_T$}}
\newcommand{\xq}{\mbox{$(x,Q^2)$}}
\newcommand{\qi}{\mbox{$q_i (x,Q^2)$}}
\newcommand{\ggl}{\mbox{$g (x,Q^2)$}}
\newcommand{\qiA}{\mbox{$q_i^{A}$}}
\newcommand{\gA}{\mbox{$g^{A}$}}
\newcommand{\dsgmdpt}{\mbox{${d\sigma}/{dp_T}$}}
\newcommand{\gp}{\mbox{$g^p$}}
\newcommand{\rhog}{\mbox{$\rho_g $}}
\newcommand{\xone}{\mbox{$x_1$}}
\newcommand{\xtwo}{\mbox{$x_2$}}
\newcommand{\qqbar}{\mbox{$q \overline{q}$}}
\newcommand{\QQbar}{\mbox{$Q \overline{Q}$}}
\newcommand{\ppbar}{\mbox{$p \overline{p}$}}
\newcommand{\jpsi}{\mbox{$J/\psi$}}
\newcommand{\xt}{\mbox{$ x_T\ $}}
\newcommand{\xbt}{\mbox{$\bar\ x_T\ $}}
\newcommand{\yj}{\mbox{$ y^{J/\psi}$}}
\newcommand{\ygam}{\mbox{$y^{\gamma}$}}

\newcommand{\be}{\begin{equation}}
\newcommand{\ee}{\end{equation}}
\newcommand{\een}{\end{subequations}}
\newcommand{\ben}{\begin{subequations}}
\newcommand{\beq}{\begin{eqalignno}}
\newcommand{\eeq}{\end{eqalignno}}
\renewcommand{\thefootnote}{\fnsymbol{footnote} }
\pagestyle{empty}
\noindent
{\flushright CERN-TH.6748/92\\}
\vspace{2cm}
\begin{center}
{\Large \bf Shedding light on quarkonium formation by\\
producing a photon in association}\\
\vspace{5mm}
{K. Sridhar$^{*)}$}\\
\vspace{8mm}
{\em Theory Division, CERN, CH-1211, Geneva 23, Switzerland}\\
\end{center}

\vspace{2.cm}
\begin{abstract}
The process $AB \rightarrow$ quarkonium $+\gamma+X$, for a heavy
quarkonium state such as $J/\psi$ or $\Upsilon$, is considered. The ratio
of the cross-sections for this process in $pp$ and \ppbar\ collisions is
shown to be a very sensitive probe of models of quarkonium formation.
\end{abstract}

\vspace{3cm}
\noindent
$^{*)} $ sridhar@vxcern.cern.ch\\

\vspace{36pt}
\noindent
CERN-TH.6748/92\\
December 1992\\
\vfill
\clearpage
\setcounter{page}{1}
\pagestyle{plain}
Over the years, there has been considerable interest in the
study of heavy quarkonia because the leptonic decay
channels of these resonances provide the cleanest signals of
heavy quark formation. The production of quarkonia takes
place through subprocesses that are dominated by gluons.
These processes have, therefore, been used to study
gluon distributions in the hadron.

The cross-section for the production of a \QQbar\ bound
state factorises into two parts~: a short-distance part, which
corresponds to the production of a heavy quark pair from a
hard collision of the incident particles, while the other part
specifies how the \QQbar\ pair produced in the collision forms
a quarkonium bound state. The short-distance part is, of
course, computable in perturbative QCD but the formation of
the bound state from the \QQbar\ pair can be specified only in
the context of some phenomenological model of hadronisation.
In particular, it is not known whether the hard scattering
subprocess is sensitive to the quantum numbers of the bound
state, or whether this information is completely screened by the
subsequent hadronisation.

For several years now, two models of quarkonium formation have
been used to describe $J/\psi$ and $\Upsilon$ production in
high energy collisions~: the colour-singlet model \cite{sing}
and the semi-local duality model \cite{dual}. The latter model
has also been referred to as the colour-evaporation model in
the literature. In the colour-singlet model, one
starts with the full \QQbar\ production amplitude and then
projects out the state with the proper spin, parity and
charge-conjugation assignments to describe a ${}^{2S+1}L_J$
quarkonium state. This state is required to be a
colour-singlet, which is achieved by the radiation of a hard
parton in the final state. The matrix element so obtained
is then convoluted with the modulus-squared of the
wave-function at the origin, $\vert R_0 \vert^2$, (or its
derivative, in the case of $P$-wave quarkonia), which
finally yields the matrix element for the quarkonium
production process. The wave-function and its derivative
are fixed from the leptonic or hadronic decay widths of
the quarkonia. In this model, because the amplitude for
the production of a state with definite $J^{PC}$ are
computed, it is possible to predict the production
cross section for different resonances in a given
family of quarkonia. It is also important to note that
the hard scattering vertex is governed by selection rules
that involve the quantum numbers of the quarkonium. For
example in leptoproduction of charmonia, the ${}^3S_1$
state (the $J/\psi$) is produced in the subprocess
$\gamma g \rightarrow {}^3S_1 g$, but the production of
the corresponding $P$-states (the $\chi$'s) is disallowed.

In the semi-local duality model, the hard scattering vertex
is completely blind to the quantum numbers of the quarkonium
resonance. The quantum numbers of the \QQbar\ pair produced
in the hard scattering are left unspecified, and the \QQbar\
pair is not required to be a colour singlet. The colour
non-singlet \QQbar\ state is assumed to form the physical
colour-singlet quarkonium state by multiple soft-gluon
emission. In practice, one simply computes the open \QQbar\
production cross-section and integrates this cross-section
between threshold ($=2m_Q$) and the open \QQbar\ production
threshold. Arguments based on semi-local duality are then
invoked to relate this integral of the open \QQbar\ production
cross-section to the sum of the resonance production
cross-sections. It is clear that the cross-section for producing
a particular resonance with a given $J^{PC}$ cannot be
predicted in this model. The cross-section for the production
of a specific bound state is a fraction $f$ of the integral
of the open-charm production cross-section over the mass
range considered. This fraction $f$ is, however, not a
prediction of the model but can be obtained by comparison
with experiment.

The colour-singlet model has been applied to the study of
quarkonium production in both lepton-hadron and hadron-hadron
collisions \cite{sing, singapp}. In leptoproduction,
the heavy-quark pair is produced through the fusion of
the incident virtual photon with a gluon from the nucleon.
The predictions of this model have been compared with the data
on $J/\psi$ production at large $p_T$ and small $z$ in
lepton-nucleon collisions \cite{singapp, nmc}, where $p_T$ is the
transverse momentum of the $J/\psi$ and $z= E_{J/\psi}/E_{\gamma}$;
$E_{J/\psi}$ and $E_{\gamma}$ are the energies of the
$J/\psi$ and the incident virtual photon, respectively.
The model gives an adequate description of the $p_T$
and rapidity distributions of the $J/\psi$ for large
$p_T$ and for $z \le$ 0.7~--~0.8. The overall normalisation
predicted by the model turns out to be lower than the
data by a factor of 2.3 \cite{nmc}. Similarly, data on $J/\psi$ and
$\Upsilon$ hadroproduction at large $p_T$ also seem to be
consistent with the predictions of this model \cite{gms}, except
for the normalisation. Comparisons of the predictions of the
semi-local duality model have also been made with data
on leptoproduction and hadroproduction of $J/\psi$ \cite{sldapp}.
The remarkable thing is that this simplistic model is also
capable of explaining the data at a level comparable to the
more elaborate colour-singlet model. A recent analysis of
$J/\psi$ production in $\mu$~--~p scattering \cite{emc}
suggests that the colour-singlet model is somewhat favoured
in the region $z \le 0.8$ as compared to the semi-local
duality model, but the experimental errors are too large for
any definite conclusions to be drawn.

In this letter, we suggest that the production of a photon in
association with the quarkonium will help to clearly distinguish
between these two models of quarkonium formation. The associated
production of $J/\psi$ and photon at large $p_T$ has been
studied in the context of collider \cite{drekim}, fixed-target
\cite{ours} and polarised $pp$ scattering experiments \cite{mine},
in the framework of the colour-singlet model; it has been shown
to be a very sensitive probe of gluon distributions. This is so
because, in the colour-singlet model, the only subprocess
that contributes to $J/\psi+\gamma$ production, with the
$J/\psi$ and $\gamma$ produced back-to-back at large $p_T$,
is the following:
\be
\label{e1}
g + g \rightarrow J/\psi + \gamma .
\ee
The colour-singlet model predicts that no $P$-states are
produced because the subprocess
\be
\label{e2}
g + g \rightarrow \chi + \gamma
\ee
is disallowed because of $C$-invariance. Moreover, the
subprocess
\be
\label{e3}
g + g \rightarrow \chi \rightarrow J/\psi + \gamma
\ee
does not produce a photon at large $p_T$ and does not
contribute to $J/\psi + \gamma$ production. Finally,
the \qqbar -initiated subprocess
\be
\label{e4}
q + \bar q \rightarrow J/\psi + \gamma
\ee
does not contribute, because the \ccbar\ thus produced will not
be in a colour-singlet state.

On the other hand, the \qqbar -initiated subprocess will
contribute in the semi-local duality model. This indicates
that these models will strongly differ in their predictions
for quarkonium$+ \gamma$ production. This difference can be
particularly enhanced if we study the ratio of the cross-sections
for quarkonium$+ \gamma$ production in $pp$ and \ppbar\
collisions. In what follows, we study the ratio
\be
\label{e5}
R \equiv {d \sigma^{pp}/dp_T \over d \sigma^{p \bar p}/dp_T} .
\ee
In the colour-singlet model, the ratio $R$ is clearly equal to
unity, because only the gluon-initiated subprocess contributes.
Moreover, this will be true for all $p_T$ and will also be
independent of the quarkonium resonance considered. Of course,
this is all true at the lowest order in QCD~: higher-order
diagrams in the colour-singlet model will have contributions
from both the gluon- and quark-initiated processes. But the
ratio $R$, which we propose to study, will be more resistant
to the effect of higher-order corrections than the absolute
cross-sections themselves. So we expect that $R$ will not
deviate significantly from unity in the colour singlet-model,
at least at fixed target energies. However, even at the
lowest order in perturbation theory, the semi-local duality
model allows for both gluon- and quark-initiated processes.
Hence, we expect that $R$ will differ from unity in this model.

We present, in the following, the $p_T$ dependence of the ratio
$R$ in the semi-local duality model for $J/\psi+\gamma$ production.
In this model, we need to integrate the open charm production
cross-section, given by the processes
\beq
\label{e6}
g + g \rightarrow c+ \bar c + \gamma , \cr
q + \bar q \rightarrow c+ \bar c + \gamma ,
\eeq
with the invariant mass of the \ccbar\ pair restricted to lie
between $4m_c^2$ and $4m_D^2$, the latter being the threshold for
open charm production. The kinematics is identical to that
described in Ref.~\cite{jpsisup}. The four-momenta of the
initial partons are taken to be
\beq
\label{e7}
&{\sqrt{s} \over 2} (1,\ \sine \theta\ \coss \phi,\ \sine \theta\ \sine
\phi,\ \coss \theta ), \cr
&{\sqrt{s} \over 2} (1,\ -\sine \theta\ \coss \phi,\ -\sine \theta\ \sine
\phi,\ -\coss \theta ).
\eeq
The $xy$ plane contains the final particles, and the direction of
the photon is taken to define the $+ve$ $x$-axis. The four-momenta
of the $c$, $\bar c$ and the photon are
\beq
\label{e8}
p_c =& {\sqrt{s} \over 2} x_3(1,\ \beta_3\ \coss \theta_{35},\
\beta_3\ \sine \theta_{35},\ 0 ), \cr
p_{\bar c} =& {\sqrt{s} \over 2} x_4(1,\ \beta_4\ \coss \theta_{45},\
\beta_4\ \sine \theta_{45},\ 0 ), \cr
p_{\gamma} =& {\sqrt{s} \over 2} x_5(1,\ 1,\ 0,\ 0 ) ,
\eeq
\be
\label{e9}
\beta_i=\sqrt{1-4m_c^2/x_i^2s}.
\ee
In the above equations, $s$ is the subprocess centre-of-mass
energy and is related to the hadronic centre-of-mass energy $S$
by the usual relation, $s=x_1x_2S$, where $x_1$ and $x_2$ are the
momentum fractions of the incident hadrons carried by the partons.
{}From energy conservation we obtain,
\be
\label{e10}
x_3+x_4+x_5=2 \hskip12pt (x_i \le 1,\ i=3,4,5).
\ee
The angles $\theta_{35}$ and $\theta_{45}$ are given by
\beq
\label{e11}
\coss \theta_{35}={(\beta_4 x_4)^2 - (\beta_3 x_3)^2 - x_5^2
\over 2 \beta_3 x_3 x_5} , \cr
\coss \theta_{45}={(\beta_3 x_3)^2 - (\beta_4 x_4)^2 - x_5^2
\over 2 \beta_4 x_4 x_5} .
\eeq
The $p_T$ of the photon (which is the same as that of the
$J/\psi$) is given by
\be
\label{e12}
p_T={\sqrt{s} \over 2} x_5 \sqrt{\coss^2\theta+\sine^2\theta
\sine^2\phi} .
\ee
The differential hadronic cross-section is then given by
\be
\label{e13}
{d\sigma \over dx_1dx_2dx_3dx_4d\phi dp_T} = {\alpha
\alpha_s^2  p_T \over 4\pi s x_5^2 \coss \theta \coss^2
\phi} H(x_1,x_2) \vert M_{q \bar q} \vert^2 + G(x_1,x_2)
\vert M_{gg} \vert^2 ,
\ee
where $H(x_1,x_2)$ and $G(x_1,x_2)$ are parton distribution
factors for the \qqbar\ and the $gg$ subprocesses, respectively.
$\vert M_{q \bar q}\vert^2$ and $\vert M_{gg}\vert^2$ are the
squared matrix elements for the subprocesses given in
(\ref{e6}). These matrix elements are obtained from those for
heavy quark leptoproduction by crossing. The matrix elements for
the latter process are given in Ref.~\cite{ellis}.

Integrating over all variables other than $p_T$ in
Eq.~(\ref{e13}), we obtained the cross-sections for $pp$
and \ppbar\ collisions as a function of $p_T$. The
integration was done by restricting the invariant mass of the
\ccbar\ pair as described above, and also restricting the
rapidities of the $J/\psi$ and the photon to be both equal
to zero. The computations are done for fixed target energies
at Fermilab ($\sqrt{s} \approx 38.75 \gev$). For the
parton densities Owens' Set~1.1 distributions \cite{owens} were
used, and the $Q^2$ for the process was taken to be $4m_c^2+p_T^2$.

The ratio $R$ of the cross-sections computed in this model
(curves labelled II)
for $J/\psi$ production is shown in Fig.~1, as a function
of $p_T$. The ratio is significantly smaller than unity and
also shows a strong $p_T$-dependence.
With increasing $p_T$, larger values of $x_1$ and
$x_2$ are sampled and the \qqbar\ process becomes more
important. Consequently, there is a larger suppression
of the ratio $R$. We have also computed $R$ for the case of
$\Upsilon$ production, and find that it also depends on
the mass of the quarkonium. In the case of $\Upsilon$, we
find that $R$ is smaller than that obtained for $J/\psi$. In
Fig.~1, we have also shown the curve (labelled I) for $R$ expected
in the colour-singlet model. The ratio is simply unity, independent
of $p_T$ and the mass of the quarkonium.

The experimental feasibility of measuring this process at
fixed-target energies was dealt with in detail in
Ref.~\cite{ours}. We would like to simply point out here
that one could get a enhancement by a factor of 100 if a nuclear
target were used and the cross-sections from $pA$ and
$\bar p A$ were compared. Nuclear effects would cancel
in the ratio.

In conclusion, we find that the production of a photon in
association with a quarkonium such as $J/\psi$ or $\Upsilon$,
will provide important insights into the nature of quarkonium
formation. In particular, by studying the ratio of the $pp$ and
\ppbar\ induced cross-sections at fixed-target energies it
will be possible to discriminate between  the colour-singlet
and the semi-local duality models of quarkonium formation.
Just as this work was being completed, we recieved a preprint
\cite{reya} where leptoproduction of $J/\psi+\gamma$ is
discussed as a probe of models of quarkonium formation.

\newpage
\section*{Figure caption}
\renewcommand{\labelenumi}{Fig. \arabic{enumi}}
\begin{enumerate}
\item   
The ratio $R$ as a function of $p_T$. I and II denote
colour-singlet and semi-local duality models, respectively.
The quarkonium states to which the curves correspond are shown
in parenthesis.
\end{enumerate}
\end{document}